\documentclass[twocolumn,aps,prb,longbibliography]{revtex4-1}
\usepackage{amsmath,amssymb,amsthm}
\usepackage{mathtools}
\usepackage{xspace}
\usepackage[usenames,dvipsnames]{xcolor}
\usepackage[colorlinks,hyperindex,breaklinks]{hyperref}
\usepackage{braket}
\usepackage{hyphenat}
\usepackage{bbold}
\usepackage{enumerate}
\usepackage{tabularx}
\usepackage[verbose]{placeins}
\usepackage[hang,flushmargin]{footmisc}
\usepackage[british]{babel}

\usepackage{srcltx}
\usepackage{color}

\usepackage{lipsum}

\newcommand{\ff}[1]{ {\boldsymbol #1} }
\newcommand{\ca}[1]{{\cal #1}}

\newcommand{\pbar}{\bar{p}}
\newcommand{\LABEL}[1]{\label{#1}}

\begin{document}
\title{One-step theory of two-photon photoemission}

\author{J. Braun$^1$, R. Rausch$^2$, M. Potthoff$^2$,  H. Ebert$^1$}

\affiliation{$^1$Department Chemie, Ludwig-Maximilians-Universit\"at M\"unchen, 81377 M\"unchen, Germany \\
$^2$I.~Institut f\"ur Theoretische Physik, Universit\"at Hamburg, 20355 Hamburg, Germany}

\begin{abstract}
A theoretical frame for two-photon photoemission is derived from the general theory of pump-probe photoemission,
assuming that not only the probe but also the pump pulse is sufficiently weak. This allows us to use a perturbative
approach to compute the lesser Green function within the Keldysh formalism. Two-photon photoemission spectroscopy
is a widely used analytical tool to study non-equilibrium phenomena in solid materials. Our theoretical approach
aims at a material-specific, realistic and quantitative description of the time-dependent spectrum based on a
picture of effectively independent electrons as described by the local-density approximation in band-structure theory.
To this end we follow Pendry's one-step theory of the photoemission process as close as possible and heavily make use
of concepts of multiple-scattering theory, such as the representation of the final state by a time-reversed low-energy
electron diffraction state. The formalism is fully relativistic and allows for a quantitative calculation of the time-dependent
photocurrent for moderately correlated systems like simple metals or more complex compounds like topological insulators.
An application to the Ag(100) surface is discussed in detail.
\end{abstract}


\maketitle

\section{Introduction}

Different pump-probe photoemission experiments have been developed in recent years as powerful techniques to study
the non-equilibrium dynamics of the electronic degrees of freedom in condensed-matter systems.
\cite{LLM+05,CMU+07,PFB+08,BPW10,RHW+11,GDP+11,GPM+11,VTT+12,CDF+12,SYA+12,RVB+12,AGJH12,WHS+12,DAB+13,MBV+13}
A widely used variant of time-resolved photoemission experiments is given by two-photon photoemission (2PPE).
\cite{AZ09,WSPD10,Fau12,RGK+14,BMA15,KRGH16} A prominent application of 2PPE is to study ultra-fast demagnetization
processes in magnetic solids. \cite{CAH+06,LLB+07,PSG+08,SPD+10} Besides the general interest in such non-equilibrium
phenomena, a variety of spectroscopic issues like dichroic phenomena, \cite{SPA+08,MPL+08,SGV+16} spin-dependent
life-times of electronic states, \cite{WSMD07,GRM+07,PSWD10} quantum beats \cite{Sch07,MSS+11} or delay-dependent
spectral-line width \cite{Sch07,MSS+11} are of great scientific interest.

While many theoretical investigations on time-dependent correlation effects in solids had been performed in recent
years, \cite{FKP09,SL13,SKM+13,USL14,KSM+14,FNF14,SCK+15,Bon16} their main focus has been on the detailed understanding
of the electronic non-equilibrium dynamics of the solid and not so much on the photoemission process itself.
Many-body calculations \cite{SL13,Bon16} are often performed for simple systems and certain simplified model
Hamiltonians, such as Hubbard-type models, to allow for an application of certain techniques such as dynamical
mean-field theory \cite{GKKR96,KSH+06,Hel07} or time-dependent density functional theory. \cite{KDE+15}

For a quantitative description of experiments, however, it is of utmost importance to additionally account for the
effects of transition-matrix elements. This is essential, for example, to address 2PPE or pump-probe experiments
performed with linearly or circularly polarized light. \cite{PSG+08,SPD+10} The corresponding polarization-dependent
effects visible in the measured spectra \cite{SPA+08,PSG+08,SPD+10} are encoded in the matrix elements. Furthermore,
the explicit consideration of the solid surface is inevitable since many pump-probe experiments are realized by
pumping into surface states \cite{WSPD10,SHR+98,Fau12,NF14,NOH+14,GMTH14} which serve as intermediate states.
A comprehensive and quantitative theory of time-resolved photoemission, including 2PPE, must therefore account
for surface-related, final-state and transition-matrix effects in a quantum-mechanically coherent and consistent
way in addition to pure modeling of the time-dependent electronic structure.

Our time-dependent one-step theory introduced recently \cite{BRP+15} represents the first step toward such a theory.
From the very beginning, it is based on the formalism of Green functions on the Keldysh contour in the complex time
plane. \cite{Kel65} This provides the natural interface with many-body theory and thereby in principle allows to
incorporate time-dependent correlation effects beyond a simple mean-field level or beyond the local-density
approximation (LDA) of band-structure theory. For the time-independent equilibrium state, it is already possible
to account for many-body correlations by combining the density-functional with dynamical mean-field theory (DFT+DMFT),
\cite{MCP+05} as has been demonstrated successfully for various correlated systems in the last years.
\cite{BME+06,SFB+09,BMM+10,MBE13} Moreover, the approach does quantitatively incorporate all matrix-element,
surface-related and final-state effects and allows us to directly compare calculated spectra with corresponding
experimental data.

The goal of the present paper is to apply the general formalism to the setup of 2PPE experiments. In this case we can
benefit from the fact that the initial pump pulse is weak and can be treated perturbatively. This greatly simplifies
the evaluation of the theory and provides us with a numerically tractable approach if, in addition, many-body
correlations beyond the LDA can be disregarded. As a proof of principle, we consider 2PPE from
the Ag(100) surface. The theory, however, is in principle applicable to a wide range of materials and problems.
Examples comprise ultrafast demagnetization processes, which are often studied for moderately correlated systems.
\cite{WSMD07,PSG+08,MSS+11,GMTH14} On the same level of accuracy our theoretical approach works for Rashba systems
or topological insulators which are recently of high scientific interest for spintronics applications.
\cite{CAH+06,NF14,RGK+14,KRGH16}

\section{One-step description of two-photon photoemission}

Angle-resolved photoemission from a system in thermal equilibrium is conventionally treated by Pendry's one-step model
\cite{Pen74,Pen76,HPT80} which describes the photoemission as a single coherent quantum process, including final-state
multiple-scattering effects in the bulk and at the surface potential, dipole selection rules and effects of the
transition-matrix elements in general by describing the photoemission final state as a time-reversed
low-energy-electron-diffraction (LEED) state. Coulomb-interaction effects are accounted for on a mean-field-like level
via the one-particle Green function in the local-density approximation of band-structure theory. \cite{HK64}
Adopting the sudden approximation implies that the description of the final state is disentangled from the
multiple-scattering theory for the initial state. The one-step model has successfully been applied to a wide range of
problems and spans photocurrent calculations ranging from a few eV to more than 10 keV \cite{GPU+11,GMU+12,MBE13a,BMK+14}
at finite temperatures and from arbitrarily ordered \cite{BMM+13} and disordered systems. \cite{BMM+10} Strong electron
correlations can be accounted for in addition via an improved many-body modeling of the initial-state Green function.
\cite{BME+06,SFB+09}

There are already a few steps toward a general theory of time-resolved photoemission by Freericks et al.,
\cite{FKP09,MDF10,SKM+13} and Eckstein et al.\ \cite{EK08,RFE16} followed by work from other groups. \cite{Ing11,USL14}
Moreover, a first realistic description of two-photon photoemission has been worked out. \cite{UG07}
The main complication for a numerically efficient computation of a time-resolved pump-probe photoemission spectrum
consists in the determination of the lesser Green function which depends on two independent time variables.
This adds to the necessity to consider realistic geometries, e.g., a semi-infinite stack of atomic layers, to
realistically model the surface region and to incorporate realistic electronic potentials typically obtained from
band-structure methods like the Korringa-Kohn-Rostoker (KKR) method. \cite{Kor47,EKM11}

To address two-photon photoemission experiments, we here start from our recently proposed ab initio theory of
pump-probe photoemission\cite{BRP+15} where the system is assumed to be exposed to a strong light pulse, described
by a light-matter interaction term $\ca V(t)$ added to the Hamiltonian $\ca H$. This drives the system's state out
of equilibrium. The pump pulse is assumed to have a finite duration. After the pump electronic relaxation processes
set in on a femtosecond time scale followed by slower relaxation mechanisms involving lattice degrees of freedom.
The latter are disregarded here. Under these conditions the following expression for the transition probability
$P_k(t)$ can be derived: \cite{FKP09,BRP+15}
\begin{eqnarray}
P_k(t)&=&\sum_{\alpha\beta} M^\ast_{k\beta} M_{k\alpha} \int_{t_0}^t dt' s_{\cal W}(t')
\int_{t_0}^t dt'' s_{\cal W}(t'')
\nonumber \\ &&
e^{-i \varepsilon(k) (t'-t'')} \langle c^\dagger_\beta(t') c_\alpha(t'') \rangle \: .
\label{eq:tdpes}
\end{eqnarray}
Here, $k$ refers to the quantum numbers of the photoelectron, $s_{\cal W}(t)$ is the envelope function for the
laser probe pulse and $M_{k\alpha}$ the photoemission matrix element. For the final state we have adopted the
sudden approximation and assumed that the Coulomb interaction of the high-energy photoelectron with the low-energy
part of the system can be neglected. The time-dependent correlation function $\langle c^\dagger_\beta(t') c_\alpha(t'')\rangle$
in Eq.\ (\ref{eq:tdpes}) can be identified with the lesser Green function $\ff G^<(t',t'')$ which depends on two
time variables and is a matrix in the orbital indices $\alpha$ and $\beta$. It is given by an equilibrium expectation
value but is time inhomogeneous as the Heisenberg time dependence is governed by the full and explicitly time-dependent
Hamiltonian $\ca H_{\rm tot}(t) = \ca H + \ca V(t)$. Accordingly, the central problem consists in computing the lesser
Green function which describes the temporal evolution of the electronic degrees of freedom after the pump pulse. 

The calculations simplify substantially when considering a system of effectively independent electrons.
In this case, the lesser Green function can be written as: \cite{BRP+15}
\begin{eqnarray}
\ff G^<(t,t') &=& \frac{1}{2} \ff G^{\rm R}(t,t_0) \int dE f_T(E) \nonumber \\ &&
(\ff G^{\rm R}_0(E)-\ff G^{\rm A}_0(E)) \ff G^{\rm A}(t_0,t') \:,
\label{eq:ggg}
\end{eqnarray}
where $f_T(E)$ denotes the Fermi distribution function and $t_0$ is a time just before the perturbation ${\cal V}(t)$
representing the pump pulse is switched on. $G^{\rm R}_0$ and $G^{\rm A}_0$ are the energy-dependent retarded and
advanced equilibrium Green functions of the unperturbed system, i.e. before the time $t_{0}$. The retarded Green
function $\ff G^{\rm R}(t,t')$ for the perturbed system must be obtained from the following integral equation
\begin{equation}
\ff G^{\rm R}(t,t') = \ff G^{\rm R}_0(t,t') + \int^t_{t'}d\tau \, \ff G^{\rm R}_0(t,\tau)
 \, {\cal V}(\tau) \,  \ff G^{\rm R}(\tau,t') \: ,
\label{eq:dysongret}
\end{equation}
where ${\cal V}(\tau)$ in its operator representation still may be an arbitrarily strong perturbation. The corresponding
advanced Green function is simply given by $\ff G^{\rm A}(t,t')=(\ff G^{\rm R}(t',t))^{\dagger}$.

To make use of these expressions we turn to the real-space representation and to a fully relativistic four-component
formulation. Eq.\ (\ref{eq:tdpes}) then reads as:
\begin{eqnarray}
P_{\ff k_{||},\varepsilon_f}(t) &=& \int d^3 r' \int d^3 r'' \int_{t_0}^t dt' \int_{t_0}^t dt''
e^{-i \varepsilon({\ff k_{||}}) (t'-t'')}
\nonumber \\ &\times&
f^{\dagger}_{\ff k_{||},\varepsilon_f}({\ff r'})~ 
{\cal W}({\ff r'},t')~G^<({{\ff r'}},t',{\ff r''},t'')~
\nonumber \\ &\times&
{\cal W}^{\dagger}({\ff r''},t'')~f_{{\ff k}_{||},\varepsilon_f}({{\ff r''}})~, \nonumber \\
\end{eqnarray}
where $\ff k_{||}$ is the component of the wave vector parallel to the surface, and $\varepsilon_f$ is the kinetic
energy of the photoelectron. Here one may use for the lesser Green function the expansion
\begin{equation}
G^{<}({\ff r},t,{\ff r'},t') = \sum_{\Lambda \Lambda'} \chi_{\Lambda}({\ \hat r})
g^{<}_{\Lambda \Lambda'}(r,t,r',t') \chi^{\dagger}_{\Lambda'}({\ \hat r'})
\label{eq:radialan}
\end{equation}
where $\chi_{\Lambda}({\ \hat r})$ denotes the relativistic spin-angular functions \cite{Bra96} with the spin-orbit
($\kappa$) and the magnetic ($\mu$) quantum numbers combined to $\Lambda$=($\kappa, \mu$). Furthermore, 
$f_{\ff k_{||},\varepsilon_f}({\ff r})$ is the single-particle-like final state of the photoelectron as usual in
the form of a time-reversed spin-polarized LEED state. The perturbation describing the probe pulse, which is assumed
as weak, is given by:
\begin{equation}
{\cal W}(\ff r, t) = {\cal W}(t) = -s_{\cal W}(t) \, e \,
 \mbox{\boldmath $\alpha$} \cdot {\ff A}_{\cal W} \: ,
\label{eq:wint}
\end{equation}
where $e$ is the electronic charge and  using the dipole approximation ${A}_{\cal W}$ denotes the spatially constant
amplitude of the electromagnetic vector potential corresponding to the radiation field and its polarisation $\lambda_{\cal W}$. 
The three components $\alpha_{k}$ of the vector $\mbox{\boldmath $\alpha$}$
are defined as the tensor product $\alpha_{k} = \sigma_{1} \otimes \sigma_{k}$ for $k=1,2,3$ and $\sigma_{k}$ denote
the 2$\times$2 Pauli spin matrices. 

With this one can finally write the photocurrent within the one-step model as:
\begin{eqnarray}
\label{eq:int}
I(\varepsilon_f,{\ff k_{||}},t)&=&\sum_{Inq \, I'n'q'}\sum_{\Lambda \Lambda'}
A^{Inq \dagger}_{\Lambda} \,
M^{Inq \, I'n'q'}_{\Lambda \Lambda' }(t) \, A^{I'n'q'}_{\Lambda'}~,
\nonumber \\
\end{eqnarray}
where $A$ represents the high-energy wave field \cite{Bra96} and where the matrix element $M$ is defined by:
\begin{eqnarray}
M^{Inq \, I'n'q'}_{\Lambda \Lambda'}(t) &=&
e^2 
\int_{t_0}^tdt' \int_{t_0}^tdt''~s_{\cal W}(t')s_{\cal W}(t'')
\nonumber \\ &&
\hspace*{-0.5cm} e^{-i\varepsilon(\ff k_{||}) (t'-t'')}   \int d^3r  \int d^3r'
\phi_{\Lambda}^{f\dagger}({\ff r})
\, \mbox{\boldmath $\alpha$} \cdot {\ff A}_{\cal W}
\, \nonumber \\ &&
\hspace*{-0.5cm} G^{< \, Inq \, I'n'q'}({\ff r},t', {\ff r}',t'')
\, \Big( \mbox{\boldmath $\alpha$} \cdot {\ff A}_{\cal W} \Big)^\dagger
\, \phi_{\Lambda'}^f({\ff r}') \; . \nonumber \\
\label{eq:mat.el}
\end{eqnarray}
Summations run over the contributions coming from atomic layers \textit{I}, atomic cells \textit{n} in layer
\textit{I} and sites \textit{q} within cell \textit{n}. 

In dealing with the perturbation ${\cal W}(t)$ (see Eq.~(\ref{eq:wint})) it is in practise often more convenient 
to use  the so-called gradient-\textit{V} form. In this case, the corresponding single electron potential
$V(\ff{r})$ at ($Inq$)  may depend in the most general case on the electronic spin. Explicit expressions for the
corresponding matrix elements split into an angular and radial part which can be found elsewhere. \cite{Bra96} 

To evaluate $G^{<}({\ff r},t,{\ff r\,'},t')$ entering Eq.~(\ref{eq:mat.el}) by use of Eq.~(\ref{eq:ggg}) together with the
Dyson equation (\ref{eq:dysongret}), the real-space representation of the retarded Green function $\ff G^{\rm R}_{0}(t,t')$
is needed. This is obtained by Fourier transformation from the energy-dependent retarded Green function
$G^{\rm R}_{0}({\ff r}, {\ff r \,'}, E)$ which in turn may be evaluated in a direct way by means of the KKR or
multiple-scattering formalism. \cite{MCVP89,EBKM16}
\begin{eqnarray}
\LABEL{EQ:GF-RH}
G^{\rm R}_{0}({\ff r}, {\ff r}', E)= -i\pbar \sum_{\Lambda} [  \, R_{\Lambda}^{Iq}({\ff r}, E) \,
H_{\Lambda}^{Iq\times}({\ff r}', E)\, \theta(r'- r)  \nonumber\\
+  H_{\Lambda}^{Iq}({\ff r}, E) \,  R_{\Lambda}^{Iq\times}({\ff r}', E) \,\theta(r - r')  ]
\delta_{II'}\delta_{nn'}\delta_{qq'}
  \nonumber\\
 +  \sum\limits_{\Lambda \Lambda '}
R_{\Lambda}^{Iq} ({\ff r}, E) \, G_{\Lambda \Lambda '}^{Inq\,I'n'q'}(E) \, R_{\Lambda '}^{I'q'\times} ({\ff r}', E)\;.
\nonumber \\
\end{eqnarray}
Here we assume that $\ff r$ is in the atomic cell ($Inq$) while $\ff r\,'$ is in ($I'n'q'$). The first term in
Eq.~(\ref{EQ:GF-RH}) is the single-site contribution to the Green function made up of the regular
($R^{Iq}_{\Lambda}$) and irregular ($H^{Iq}_{\Lambda}$) solutions to the single site Dirac equation for site ($Inq$)
where the index \textit{n} can be dropped because of the two-dimensional periodicity in a layer $I$. The sign
"$\times$" distinguishes left-hand-side solutions for the Dirac equation from the standard right-hand-side ones.
\cite{EBKM16} The second term in Eq.~(\ref{EQ:GF-RH}) represents the back-scattering term which accounts for 
all scattering events between sites ($Inq$) and ($I'n'q'$) in a self-consistent way. \cite{EBKM16} To calculate the
so-called structural Green function $G^{Inq\,I'n'q'}_{\Lambda\Lambda'}$ occurring in that term several techniques
are available. \cite{EKM11,MCVP89} Finally, the energy-dependent factor $\pbar$ represents essentially the relativistic
momentum. \cite{EBKM16}

In the following an application of the scheme to 2PPE spectroscopy will be presented. Accordingly, we now
assume that both, the pump ${\cal V}(\tau)$ and the probe pulse ${\cal W}(\tau)$, are weak in intensity.
As mentioned before, this situation quantitatively describes the scenario of two-photon photoemission
spectroscopy. Consequently, Eq.~(\ref{eq:dysongret}) can be solved perturbatively in first-order approximation by
replacing $\ff G^{\rm R}$ by $\ff G^{\rm R}_0$ on the right side of Eq.~(\ref{eq:dysongret}). This leaves us with
a simple integral expression for $\ff G^{\rm R}$ while $\ff G^{\rm R}_0$ is available from Eq.~(\ref{EQ:GF-RH}).

As a first step, we calculate the atomic-like contribution using the single-site part of Eq.~(\ref{EQ:GF-RH}) only.
Substituting the Fourier transform of the single-scattering contribution into Eq.~(\ref{eq:dysongret}), this
site-diagonal term is obtained as:
\begin{widetext}
\begin{eqnarray}
\LABEL{EQ:GR-AT}
G^{{\rm R}\, Inq\,Inq}_{\rm at}({\ff r},t,{\ff r}',t') &=&
 -\frac{i\pbar}{2\pi} \sum_{\Lambda} \int dE e^{-iE(t-t')} \,
 \, \Big[ R^{Iq}_{\Lambda}({\ff r}, E) \, H_{\Lambda}^{Iq\times}({\ff r}', E)\, \theta(r'- r)
+  H^{Iq}_{\Lambda}({\ff r}, E) \,  R_{\Lambda}^{Iq\times}({\ff r}', E) \,\theta(r - r') \Big]
\nonumber \\ &-&
 \frac{\pbar^2}{4\pi^2} \sum_{\Lambda\Lambda'} \int^t_{t'}dt''s_{\cal V}(t'') \int dE e^{-iE(t-t'')} \int dE' e^{-iE'(t''-t')}
\nonumber \\ &\times& \;
\Big[  H^{Iq}_{\Lambda}({\ff r},E) \,
M^{(1)\, Iq}_{\Lambda\Lambda'}(r',r,E,E') \, R^{Iq\times}_{\Lambda'}({\ff r}',E')
+ 
       H^{Iq}_{\Lambda}({\ff r},E) \,
M^{(2)\, Iq}_{\Lambda\Lambda'}(0,r_<,E,E') \, H^{Iq\times}_{\Lambda'}({\ff r}',E')
\nonumber \\ &&
+R^{Iq}_{\Lambda}({\ff r},E) \,
M^{(3)\, Iq}_{\Lambda\Lambda'}(r_>,r_{\rm cr},E,E') \, R^{Iq\times}_{\Lambda'}({\ff r}',E')
      +R^{Iq}_{\Lambda}({\ff r},E) \,
M^{(4)\, Iq}_{\Lambda\Lambda'}(r,r',E,E') \, H^{Iq\times}_{\Lambda'}({\ff r}',E')
\Big] \;,  \nonumber \\
\end{eqnarray}
with the four matrix element functions
\begin{eqnarray}
\LABEL{EQ:MATFUN1}
M^{(1)\, Iq}_{\Lambda\Lambda'}(r_a,r_b,E,E') &=&
\int^{r_b}_{r_a}d^3r''
R^{Iq\times}_{\Lambda}({\ff r}'',E)
\, e \mbox{\boldmath $\alpha$} \cdot {\ff A}_{\cal V}  \,
 H^{Iq}_{\Lambda'}({\ff r}'',E') \: , 
\\
M^{(2)\, Iq}_{\Lambda\Lambda'}(r_a,r_b,E,E')&=&
\int^{r_b}_{r_a}d^3r'' R^{Iq\times}_{\Lambda}({\ff r}'',E)
\, e \mbox{\boldmath $\alpha$} \cdot {\ff A}_{\cal V}  \,
 R^{Iq}_{\Lambda'}({\ff r}'',E')\: , 
 \\
M^{(3)\, Iq}_{\Lambda\Lambda'}(r_a,r_b,E,E') &=&
\int_{r_a}^{r_b}d^3r'' H^{Iq\times}_{\Lambda}({\ff r}'',E)
\, e \mbox{\boldmath $\alpha$} \cdot {\ff A}_{\cal V}  
 \, H^{Iq}_{\Lambda'}({\ff r}'',E')\: , 
 \\
M^{(4)\, Iq}_{\Lambda\Lambda'}(r_a,r_b,E,E')&=&
\int_{r_a}^{r_b}d^3r'' H^{Iq\times}_{\Lambda}({\ff r}'',E)
\, e \mbox{\boldmath $\alpha$} \cdot {\ff A}_{\cal V}  \,
 R^{Iq}_{\Lambda'}({\ff r}'',E') \; .
\LABEL{EQ:MATFUN4}
\end{eqnarray}

Here the real space representation for the pump pulse ${\cal V}(\tau)$ has been split in analogy to 
Eq.\ (\ref{eq:wint}) for the probe pulse  ${\cal W}(\tau)$ and $r_{\rm cr}$ is the critical radius
of the sphere bounding the atomic cell and use has been made of the fact that the system has two-dimensional
periodicity. The symbols $r_<$ and $r_>$ in Eq.~(\ref{EQ:GR-AT}) are defined as $r_< = min(r,r')$
and $r_> = max(r,r')$. In  Eq.~(\ref{EQ:MATFUN1})-(\ref{EQ:MATFUN4}) $r_{\rm a}$ and $r_{\rm b}$ serve as
dummy variables for the corresponding integration boundaries. Furthermore, for $M^{(1)}$ and $M^{(4)}$ the
constraint $r_{\rm b} > r_{\rm a}$ must hold, otherwise they are zero.
In addition to this atomic-like part, the retarded Green function leads to two mixed contributions
between single-scattering and multiple-scattering events, as well as a double multiple-scattering contribution.
First we present the two mixed contributions. The first one is given by: 
\begin{eqnarray}
G^{{\rm R}\, Inq\,I'n'q'}_{m1}({\ff r},t,{\ff r}',t') &=&
-\frac{i\pbar}{4\pi^2} \sum_{\Lambda \Lambda' \Lambda''} \int^t_{t'}dt''s_{\cal V}(t'')
\int dE e^{-iE(t-t'')}\int dE' e^{-iE'(t''-t')}
\nonumber \\ &\times&
      \Big[ H^{Iq}_{\Lambda}({\ff r},E) \,
M^{(2)\, Iq}_{\Lambda\Lambda'}(0,r,E,E')
 \, G_{\Lambda' \Lambda ''}^{Inq\,I'n'q'}(E') \,
R^{I'q'\times}_{\Lambda''}({\ff r}',E')
\nonumber \\ &&
+R^{Iq}_{\Lambda}({\ff r},E)
M^{(4)\, Iq}_{\Lambda\Lambda'}(r,r_{\rm cr},E,E')
 \, G_{\Lambda' \Lambda ''}^{Inq\,I'n'q'}(E') \,
R^{I'q'\times}_{\Lambda''}({\ff r}',E') \Big]~.
\end{eqnarray}
Here the structural Green function $G^{Inq\,I'n'q'}_{\Lambda\Lambda'}$ accounts for all multiple-scattering
events for the propagation from site $(I'n'q')$ to  site $(Inq)$.

For the second mixed contribution we find:
\begin{eqnarray}
G^{{\rm R}\, Inq\,I'n'q'}_{m2}({\ff r},t,{\ff r}',t') &=&
-\frac{i\pbar}{4\pi^2} \sum_{\Lambda \Lambda'  \Lambda''} \int^t_{t'}dt''s_{\cal V}(t'')
\int dE e^{-iE(t-t'')} \int dE' e^{-iE'(t''-t')}
\nonumber \\ &\times&
      \Big[ R^{Iq}_{\Lambda}({\ff r},E)
 \, G_{\Lambda \Lambda '}^{Inq\,I'n'q'}(E) \,
M^{(1) \, I'q'}_{\Lambda'\Lambda''}(r',r_{\rm cr},E,E') \,
R^{I'q'\times}_{\Lambda''}({\ff r}',E')
\nonumber \\ &&
+R^{Iq}_{\Lambda}({\ff r},E)
 \, G_{\Lambda \Lambda '}^{Inq\,I'n'q'}(E)
M^{(2) \, I'q'}_{\Lambda'\Lambda''}(0,r',E,E') \,
         \, H^{I'q'\times}_{\Lambda''}({\ff r}',E') \Big]~.
\end{eqnarray}
As the last bulk-like contribution the multiple-to-multiple-scattering part
$G^{{\rm R}\, Inq \,I'n'q'}_{mm}$ is obtained as
\begin{eqnarray}
G^{{\rm R}\, Inq \,I'n'q'}_{mm}({\ff r},t,{\ff r}',t')
&=& \frac{1}{2\pi}\sum_{\Lambda \Lambda '} \int dE e^{-iE(t-t')}
R^{Iq}_{\Lambda}({\ff r},E)
 \, G_{\Lambda \Lambda '}^{Inq \,Inq'}(E) \,
R^{I'q'\times}_{\Lambda'}({\ff r}',E)
\nonumber \\ &+&
\frac{1}{4\pi^2} \sum_{I''n''q''} \sum_{\Lambda \Lambda '} \sum_{\Lambda'' \Lambda'''}\int^t_{t'}dt''s_{\cal V}(t'')
\int dE e^{-iE(t-t'')} \int dE' e^{-iE'(t''-t')}
\nonumber \\ &\times&
         R^{Iq}_{\Lambda}({\ff r},E)
      \, G_{\Lambda \Lambda '}^{Inq \,I''n''q''}(E)
      \, M^{(2)\, I''q''}_{\Lambda'\Lambda''}(0,r_{\rm cr},E,E')
      \, G_{\Lambda'' \Lambda '''}^{I''n''q''\,I'n'q'}(E') \,
         R^{I'n'q'\times}_{\Lambda'''}({\ff r}',E') \;~. \nonumber \\ 
\end{eqnarray}
\end{widetext}
The total double-time dependent bulk-like radial part of the retarded Green function follows by
summing the four different contributions presented above:
\begin{eqnarray}
G^{{\rm R} \, Inq \,I'n'q'}({\ff r},t,{\ff r}',t') &=& G^{{\rm R} \, InqInq}_{at}({\ff r},t,{\ff r}',t')
\nonumber \\ &+&
G^{{\rm R} \, Inq \, I'n'q'}_{m1}({\ff r},t,{\ff r}',t')
\nonumber \\ &+& G^{{\rm R} \, Inq \, I'n'q'}_{m2}({\ff r},t,{\ff r}',t')
\nonumber \\ &+&
G^{{\rm R} \, Inq \,I'n'q'}_{mm}({\ff r},t,{\ff r}',t')~. \nonumber \\
\end{eqnarray}

It remains to calculate the surface contribution of the time-resolved photocurrent. Especially
in the case where the image-potential states \cite{GBB+93} serve as intermediate states in
the two-photon photoemission process, it is essential to describe all surface-related electronic
states including the image states in a realistic way. In the static version of the one-step model
this is usually realized by employing a Rundgren-Malmstr\"om barrier potential. \cite{MR80} This approach
can be generalized to the time-dependent case. We start with an appropriate formulation of the retarded
free-electron Green function
\begin{eqnarray}
G^{\rm R}_{0 ,\rm surf}({\ff r},{\ff r'},E)~=~\sum_{{\ff k}_{\ff g}}~\frac{\Psi_{{\ff k}_{\ff g}}({\ff r})
\Psi^{\dagger}_{{\ff k}_{\ff g}}({\ff r'})} {E-\varepsilon({\ff k}_{\ff g})+i\delta}~,
\end{eqnarray}
where ${\ff g}$ denotes a two-dimensional reciprocal lattice vector and ${\ff k}_{\ff g}$ the corresponding
wave vector for energy $\varepsilon=\varepsilon({\ff k}_{\ff g})$. The Fourier transform is
\begin{eqnarray}
G^{\rm R}_{0 , \rm surf}({\ff r},t,{\ff r'},t')&=&-i\theta(t-t')\sum_{{\ff k}_{\ff g}}\Psi_{{\ff k}_{\ff g}}({\ff r})
\Psi^{\dagger}_{{\ff k}_{\ff g}}({\ff r'})
\nonumber \\ && e^{-i\varepsilon({\ff k}_{\ff g})(t-t')}~.
\end{eqnarray}
The scattering properties of the surface potential can be expressed in terms of the barrier scattering matrix
$M^{\pm \pm}_{{\ff g}{\ff g}}\delta_{{\ff g}{\ff g'}}$. With the Kronecker delta, $M$ is a diagonal matrix, i.e.,
corrugation effects in the surface potential are neglected. \cite{MR80} This matrix is represented as an additional
surface layer in the formalism and is typically located in front of the first atomic layer at the distance
$z_{\rm S}$, which defines the so-called image plane. \cite{MR80} Given this matrix, plane-wave amplitudes
$a^+_{0{\ff g}}$ and $d^-_{0{\ff g}}$ can be defined, where the first amplitude is emitted by the surface barrier
and the second one is emitted by the semi-infinite stack of atomic layers. \cite{HPT80,Bra96} The corresponding
amplitude emitted from the semi-infinite bulk may be denoted by $b^-_{1{\ff g}}$, and $d^+_{1{\ff g}}$ represents
its reflected counterpart. These four amplitudes satisfy the following linear system of equations:
\begin{eqnarray}
d^-_{0{\ff g}}~=~P^-_{0{\ff g}}b^-_{1{\ff g}}~+~P^-_{0{\ff g}}\sum_{\ff g'}R^{-+}_{{\ff g}{\ff g'}}d^+_{1{\ff g'}}~,
\nonumber \\
d^+_{1{\ff g}}~=~P^+_{0{\ff g}}a^+_{0{\ff g}}~+~P^+_{0{\ff g}}\sum_{\ff g'}M^{+-}_{{\ff g}{\ff g \,'}}d^-_{0{\ff g'}}~,
\end{eqnarray}
where $P^{\pm}$ denote the free electron Green functions which propagate the wave field between the first
atomic layer and the surface layer. The amplitude $a^+_{0{\ff g}}$ can also be calculated by standard 
multiple-scattering techniques. The result is:
\begin{eqnarray}
        a^+_{0{\ff g}}(z_{\rm S})=
        \frac{A_z e^{-i{\ff q}\cdot{\ff c}_S}}{2\omega c k_{{\ff g}z_{\rm S}}^+}
\hspace*{-0.2cm}\int\limits_{-\infty}^{z_{\rm S}}\hspace*{-0.2cm}dze^{-i(q_z+k_{{\ff g}z_{\rm S}}^+)(z-c_{z_{\rm S}})}
        V'_{\rm B}\psi_{{\ff g}}^*(z)~. \nonumber \\
\end{eqnarray}
This plane wave amplitude appears directly proportional to the gradient of the surface potential $V_{\rm B}$.
Here, $R^{-+}$ denotes the bulk-reflection matrix and $\psi_{{\ff g}}^*(z)$ represents the wave function
of the intermediate state in the barrier region. The amplitude $b^-_{1{\ff g}}$ is also known from standard
multiple-scattering techniques applied to the calculation of the initial-state wave function between the
different layers of the semi-infinite bulk. \cite{HPT80} Consequently, the two amplitudes $d^-_{0{\ff g}}$
and $d^+_{1{\ff g}}$ can be calculated from the system of linear equations which is defined above.
For $d^-_{0{\ff g}}$ we find:
\begin{eqnarray}
&&d^{-}_{0{\ff g}} = \sum_{\ff g'} \left\{
(1-P^{-}R^{-+}P^{+}M^{+-})^{-1}P^{-}\right\}_{ {\ff g}{\ff g'}} b^{-}_{1{\ff g}'}
\nonumber\\
&+& \sum_{\ff g \,'} \left\{(1-P^{-}R^{-+}P^{+}M^{+-})^{-1}P^{-}R^{-+}P^{+} \right\}_{{\ff g} {\ff g}'}
a^{+}_{0{\ff g}'}~, \nonumber \\
\end{eqnarray}
and $d^+_{1{\ff g}}$ results in:
\begin{eqnarray}
d^{+}_{1{\ff g}} = \sum_{\ff g'} (P^{+}M^{+-})_{ {\ff g}{\ff g'}} d^{-}_{0{\ff g'}}  +
\sum_{\ff g'} P^{+}_{ {\ff g}{\ff g \,'}} a^{+}_{0{\ff g'}}~.
\end{eqnarray}
Having these coefficients calculated, the wave field of the intermediate state at the image plane can be
expressed within a plane-wave representation:
\begin{eqnarray}
\Psi_{{\ff k}_{\ff g}}({\ff r})~=~\left(a^{+}_{0{\ff g}}e^{i{\ff k}^+_{\ff g}({\ff r}-{\ff r}_{S})}+
d^{-}_{0{\ff g}} e^{i{\ff k}^-_{\ff g}({\ff r}-{\ff r}_{S})} \right)~. \nonumber \\
\end{eqnarray}
The corresponding expansion in spherical harmonics gives:
\begin{eqnarray}
\Psi_{{\ff k}_{\ff g}}({\ff r})~=~\sum_{\Lambda}H_{\Lambda}\Psi_{\Lambda,{\ff k}_{\ff g}}(r)
\chi_{\Lambda}({\hat r})~,
\end{eqnarray}
with the spherical coefficients
\begin{eqnarray}
H_{\Lambda'}&=&\sum_{{\ff g}s} 4\pi i^{l'}(-2s)(-)^{\mu' -s} C^{\Lambda'}_{s}
\nonumber \\ &&
\big[ a^{+}_{0{\ff g}s} Y^{s-\mu'}_{l'} (\widehat{k^{+}_{{\ff g}}}) + d^{-}_{0{\ff g}s} Y^{s-\mu'}_{l'}
(\widehat{k^{-}_{{\ff g}}}) \big]~.
\end{eqnarray}
For an explicit calculation of the retarded surface Green function the radial part of the spherical wave
field is needed. Unfortunately, the surface potential is a function of the Cartesian coordinate $z$ only,
and so is the corresponding wave function. This means that the wave function is not directly available. Nevertheless,
as a good approximation we define:
\begin{eqnarray}
\Psi_{\Lambda,{\ff k}_{\ff g}}(z) \approx \Psi^{\rm LK}_{\Lambda,{\ff k}_{\ff g}}(r)~,
\end{eqnarray}
where $\Psi^{\rm LK}_{\Lambda,{\ff k}_{\ff g}}(r)$ represents the spherical wave function which belongs to an
empty-sphere potential of the first vacuum layer. This approximation works well because in self-consistent
TB-SPRKKR \cite{EKM11} electronic-structure calculation a set of vacuum layers is typically used, which is
located on top of the first atomic layer. Due to the charge transfer from the first atomic layer into the
empty sphere potentials of the first and second vacuum layers, these empty-sphere potentials represent in a
reasonable approximation the polynomial region of the surface potential. 
Within the local spin-density formalism, the surface retarded Green function is then given by
\begin{eqnarray}
&&G^{\rm R}_{0 ,\rm surf}({\ff r},t,{\ff r}',t')=\hspace*{-0.06cm} \sum_{\Lambda,\Lambda'{\ff k}_{\ff g}}\hspace*{-0.12cm}
e^{-i\varepsilon({\ff k}_{\ff g})(t-t')} \nonumber \\ &&
H_{\Lambda} H^*_{\Lambda'} \Psi^{\rm LK}_{\Lambda,{\ff k}_{\ff g}}(r)\Psi^{{\rm LK} \dagger}_{\Lambda',{\ff k}_{\ff g}}(r)
\chi_{\Lambda}({\hat {\ff r}})\chi^{\dagger}_{\Lambda'}({\hat {\ff r'}})~,
\end{eqnarray}
where the $\theta$-function has been omitted since the condition $t>t'$ is fulfilled. Inserting the expression
for $G^{\rm R}_{0, \rm surf}$ in Eq.~(\ref{eq:dysongret}), the retarded surface Green function in first-order
approximation reads
\begin{eqnarray}
G^{\rm R}_{\rm surf}({\ff r},t,{\ff r}',t') &=& \sum_{\Lambda,\Lambda'{\ff k}_{\ff g}{\ff k}_{\ff g'}}
e^{-i\varepsilon({\ff k}_{\ff g})(t-t')}
\nonumber \\ &&
E_{{\ff k}_{\ff g}{\ff k}_{\ff g'}}(t,t') \Psi^{\rm LK}_{\Lambda,{\ff k}_{\ff g}}({\ff r})
\Psi^{{\rm LK} \dagger}_{\Lambda',{\ff k}_{\ff g'}}({\ff r}')
\nonumber \\ &&
\end{eqnarray}
with
\begin{eqnarray}
E_{{\ff k}_{\ff g}{\ff k}_{\ff g \,'}}(t,t') &=& \delta_{{\ff k}_{\ff g}{\ff k}_{\ff g'}}-
B_{{\ff k}_{\ff g}{\ff k}_{\ff g'}} F_{{\ff k}_{\ff g}{\ff k}_{\ff g'}}(t,t')~,
\end{eqnarray}
and where
\begin{eqnarray}
F_{{\ff k}_{\ff g}{\ff k}_{\ff g'}}(t,t') &=& \int_{t'}^{t} dt'' s_{\nu}(t'')
e^{-i(\varepsilon_{{\ff k}_{\ff g'}}-\varepsilon_{{\ff k}_{\ff g}})(t''-t')}
\end{eqnarray}
and
\begin{eqnarray}
B_{{\ff k}_{\ff g}{\ff k}_{\ff g'}} = \sum_{\Lambda \Lambda'}
H^*_{\Lambda}M^{\rm surf}_{\begin{substack}{ \Lambda\Lambda'\\
{\ff k}_{\ff g}{\ff k}_{\ff g'}} \end{substack}} H_{\Lambda'}~.
\end{eqnarray}
The radial part of the surface matrix element is defined as:
\begin{eqnarray}
M^{\rm surf}_{\begin{substack}{ \Lambda\Lambda'\\
{\ff k}_{\ff g}{\ff k}_{\ff g'}} \end{substack}}
= \int_{0}^{r_{\rm cr}} dr''{^3} \Psi^{\times}_{\Lambda{\ff k}_{\ff g}}({\ff r}'')
\, e \mbox{\boldmath $\alpha$} \cdot {\ff A}_{\cal V}  \,
 \Psi_{\Lambda'{\ff k}_{\ff g'}}({\ff r}'')~. \nonumber \\
\end{eqnarray}
The total double-time-dependent radial part of the retarded Green function follows by summing the five
contributions discussed above:
\begin{eqnarray}
G^{{\rm R} \, Inq \, I'n'q'}({\ff r},t,{\ff r}',t') &=& G^{{\rm R} \, InIn}_{at}({\ff r},t,{\ff r}',t')
\nonumber \\ &+&
G^{{\rm R} \, Inq \, I'n'q'}_{m1}({\ff r},t,{\ff r}',t')
\nonumber \\ &+& G^{{\rm R} \, Inq \, I'n'q'}_{m2}({\ff r},t,{\ff r}',t')
\nonumber \\ &+&
G^{{\rm R} \, Inq \, I'n'q'}_{mm}({\ff r},t,{\ff r}',t')~
\nonumber \\ &+&
G^{\rm R}_{\rm surf}({\ff r},t,{\ff r}',t')~.
\end{eqnarray}
Having $G^{{\rm R} \, Inq \, I'n'q'}$ calculated numerically, $G^{< \, Inq \, I'n'q'}$ can be obtained via
Eq.~(\ref{eq:ggg}). Finally, with Eqs.\ (\ref{eq:int}) and (\ref{eq:mat.el}) one may compute the time-dependent
2PPE signal.
 
\section{2PPE from Ag(100)}

\begin{figure}[b]
\includegraphics[width=0.4\textwidth,clip]{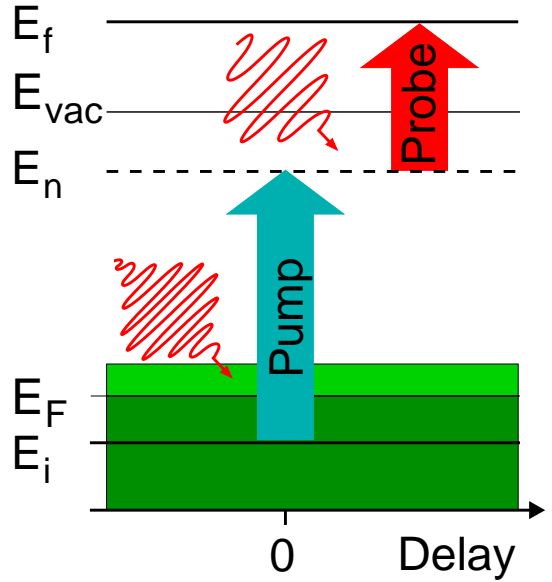}
\label{FIG-scheme}
\caption {(Color online) 
Schematic presentation of 2PPE process with the energy-resolved measurement mode for
spin-integrated spectroscopy.}
\end{figure}
As a first application the theory is applied to the (100) surface of Ag, i.e., to a prototypical
simple paramagnetic metal. We compute the lesser Green function and the 2PPE spectrum within
multiple-scattering theory in a fully relativistic way by using the Munich SPRKKR program package
in its tight-binding version.\cite{SPR-KKR6.3} The spherically symmetric potential was obtained
within atomic-sphere approximation (ASA) and the corresponding single-site wave functions serve
as input quantities for the calculation of the lesser Green function. The latter is obtained in two
steps: First we determine the retarded Green function from the respective Dyson equation (\ref{eq:dysongret})
where we treat the perturbation to first order in the pump-pulse strength. Second, the lesser Green
function is calculated from Eq.~(\ref{eq:ggg}). The solutions of the conditional equations for the Green
functions are obtained by numerical matrix operations, where the expansion into spherical harmonics
includes orbital quantum numbers up to $l=2$. The two radial coordinates $r$ and $r'$ are restricted
to the ASA sphere. With respect to the dynamical degrees of freedom, the equations are Fourier-transformed
from time to energy space. To this end we choose an equidistant mesh for the two time variables $t$ and
$t'$ with a time step of $\Delta t=1$~fs. The energy-dependent retarded KKR Green function is calculated
for a complex energy, with an energy-dependent imaginary part $V_i(E)=0.05 + 0.01 (E-E_{\rm F})^2$ in {\rm eV},
to account for damping effects due to inelastic scattering events. Therewith, the life-time broadenings
of the first and second images states, $\Gamma_1=21$ meV and $\Gamma_2=5$ meV, are accounted for
quantitatively. \cite{SFFS92} Concerning the numerical effort, this also helps to reduce the number of
energy grid points necessary for a numerically accurate Fourier transformation. Converged results are
obtained for an energy window of about 3.0~eV around the Fermi level $E_{\rm F} \equiv 0$.

\begin{figure}[t]
\includegraphics[width=0.45\textwidth,clip]{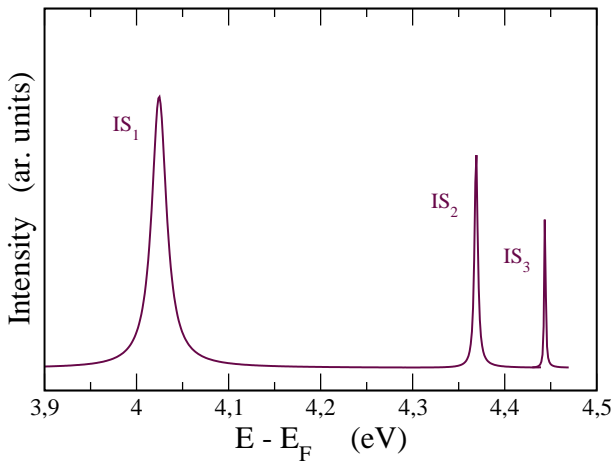}
\caption {(Color online)
Intensity profiles of the first three Rydberg states calculated for the Ag(100) surface with an
excitation energy of 6.02~eV. The work function has been set to $\Phi=4.6$~eV.}
\end{figure}
Our calculations refer to the energy-resolved operational mode of an 2PPE experiment as schematically
shown in Fig.~1. \cite{Pic07} In this mode the time delay between the pump and probe pulses is fixed, and a
kinetic-energy spectrum of the excited electrons is recorded. For a given kinetic energy $E_{\rm kin}$,
the energy of the intermediate state $E$, with respect to the Fermi level $E_{\rm F}=0$, is obtained by
$E=E_{\rm kin}-h\nu_{\rm probe}+\Phi$ from the photon energy of the probe pulse $h\nu_{\rm probe}$ and
the work function $\Phi$. The initial-state energy is $E_i = E_{\rm kin}-h\nu_{\rm pump}-h\nu_{\rm probe}+\Phi$.
$E_{\rm vac}$ denotes the vacuum level in Fig.~1.

Calculations are performed for a Gaussian pump pulse with mean energy $h\nu_{\rm pump} = 4.02$~eV to reach
the first image-potential state of the Ag(100) surface which is located at about 4~eV above the Fermi level
(the work function is $\Phi=4.6$~eV). The full width at half maximum of the pulse is chosen as FWHM $= 2.0$~fs,
and the maximum amplitude of the pulse is located at time $t=3.0$~fs, i.e., for $t_0=0$ [see Eqs.\ (\ref{eq:ggg})
and (\ref{eq:dysongret})] the system can be assumed to be in its ground state. In principle, this setup provides
the possibility to study the dependence of energetics and dynamics on the parallel component of the wave vector
$\ff k_{||}$, but we restrict the present study to normal emission, $\ff k_{||}=0$.
\begin{figure}[b]
\includegraphics[width=0.48\textwidth,clip]{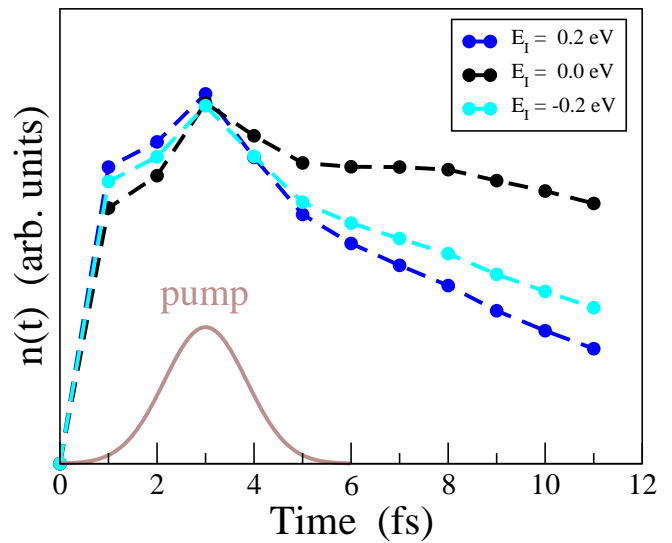}
\caption {(Color online)
Time-dependent occupation number $n(t)$ of the first image state calculated for the Ag(100) surface as
a function of the initial-state energy $E_i$.}
\end{figure}
Using a pump photon energy of $h\nu_{\rm pump}=4.02$~eV it is possible to directly excite the first image state
which is considered to serve as intermediate state in our pump-probe scenario. 

In Fig.~2 we present a conventional inverse photoemission spectrum calculated in normal emission with linear p-polarized light. 
The spectrum uncovers the first few image-potential states at energies just below the vacuum level (4.6~eV above $E_{\rm F}$). 
As the most prominent feature we observe a sharp peak at 4.02~eV above $E_{\rm F}$, which is identified as the first
image state of the Ag(100) surface. The second and the third image states are also visible and show up at about
4.37~eV and 4.44~eV above $E_{\rm F}$, respectively.

Before proceeding with the discussion to the actual 2PPE calculations, we concentrate on the time-dependent
population $n(t)$ of the intermediate image state. This can be computed from the lesser Green function by
integrating over a representative sphere within the surface layer
\begin{eqnarray}
n(t) = - i \int_{\rm surf} d^3r~G^{<}({\ff r},t,{\ff r},t)~.
\end{eqnarray}
The use of a single ASA sphere which is located in the first atomic layer is well justified because an image state
is represented by a two dimensional electron gas located in front of the first atomic layer, pinned in energy just
below the vacuum level. $n(t)$ is shown in Fig.~3 for different initial-state energies $E_i$ around $E_{\rm F}=0$.
As expected, the population of the first image-potential state increases with time, where the maximum population
appears at the maximum amplitude of the pump pulse, and then decreases again. Physically, the depopulation of the
image state on a femtosecond time scale is mediated by electron-electron scattering processes which, however, are
not yet included in the formalism explicitly. These processes are rather accounted for on a phenomenological level
by the energy-dependent life-time, i.e., by the imaginary part of the complex energy. Furthermore, we find that the
depopulation of the first image state sensitively depends on the initial-state energy. At $E_i=0$ the relaxation
is much slower and reveals a plateau-like structure for later times while for $E_{i}=\pm 0.2$~eV the decay of
$n(t)$ is faster. This can be ascribed to the fact that, for the given pump energy, only at $E_{i}=0$ there are
physical excitations from occupied states at $E_{\rm F}$ into the first image state while for $E_{i}\ne 0$
transitions are virtual, i.e., they more or less violate the energy-conservation condition and must decay exponentially
on a much shorter time scale.
\begin{figure}[t]
\includegraphics[width=0.48\textwidth,clip]{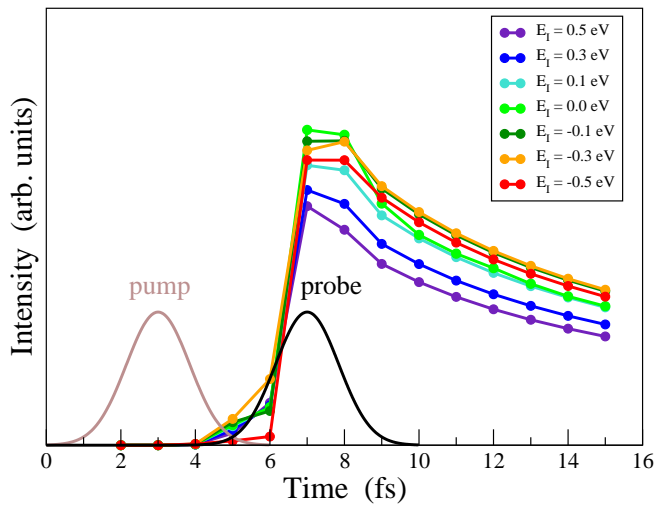}
\caption {(Color online)
Time-dependent intensity distributions $P(t)$ as a function of the initial-state energy $E_i$ for linear p-polarized light.}
\end{figure}

In Fig.~4 we present a series of calculated time-dependent total intensities in normal emission. Time-dependent
photoemission intensities $P(t)$ are shown for linearly p-polarized light and for different initial-state energies
$E_i$ ranging from $-0.5$~eV to $0.5$~eV relative to $E_{\rm F}=0$. The calculations have been done for the same
pump as discussed above and for a probe pulse with a mean energy of 2.0~eV to lift the electrons from the intermediate
image state to states above the vacuum level. If energy conservation held, i.e., disregarding the uncertainty principle,
their kinetic energy would be given by 1.4~eV. The maximum amplitude of the Gaussian profiles of the pump and probe
pulses, with FWHM 2.0~fs each, have been placed at 3.0~fs and 7.0~fs, respectively, i.e., the time delay is fixed at 4.0~fs.

As can be seen in Fig.~4, the maximum intensity is reached for all initial-state energies at about 7.0~fs, i.e., at the
temporal peak maximum of the probe pulse -- as expected. As a function of the kinetic energy, or equivalently as a
function of $E_{i}$, and starting from $E_i=-0.5$~eV, the intensity increases, reaches its maximum for $E_i=0$ and then
decreases again. This behavior reflects the peak-like structure in the corresponding 2PPE experiment with a peak maximum
at about 1.4 eV kinetic energy.
\begin{figure}[b]
\includegraphics[width=0.42\textwidth,clip]{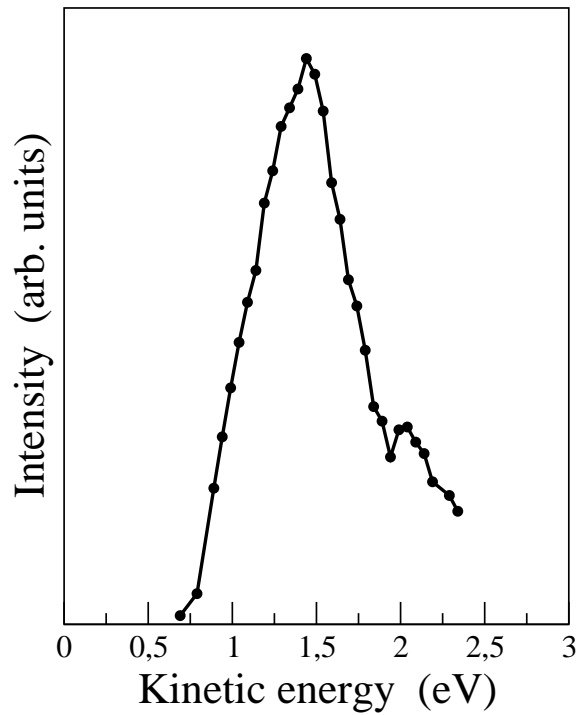}
\caption {(Color online)
Integrated intensity rates calculated for the energy-resolved mode for a fixed pump-probe delay of 4 fs (see Fig.4).
Each point in the intensity distribution represents a time-dependent intensity rate integrated over the interval
from 0--15 fs. The corresponding time-dependent intensity rates were taken from Fig.~4.}
\end{figure}

However, the spectra presented in Fig.~4 cannot be compared directly with corresponding experimental data, because a
2PPE experiment usually works in a time-integrated mode. The time resolution is defined by the pump-probe time delay and
typically lies in a regime of 100 fs up to 1 ps.\cite{WSPD10,WSMD07} While the time scale considered in our present model
calculation is much shorter, this should not be seen as a major problem. As long as the first-order approximation in
Eq.\ (\ref{eq:dysongret}) is valid, the time scales are essentially controlled by the retarded KKR Green function which is
homogeneous in time, $\ff G_{0}^{\rm R}(t,t')=\ff G_{0}^{\rm R}(t-t')$. Hence, substantially longer times are accessible
with essentially the same computational effort. Only the number of energy values $E$ and $E'$, which are used in the
Fourier transformations must be increased by a certain factor to guarantee for a comparable numerical accuracy in the
calculation of $P(t)$.

In principle, comparison with the experimental 2PPE spectrum can be achieved by integrating the time-dependent
intensities shown in Fig.~4 for each $E_{i}$. Here, this is done for the time interval up to $t=15$~fs. Fig.~5
displays the resulting 2PPE spectra where the time-averaged intensity is plotted as a function of the kinetic energy
corresponding to the respective initial-state energy $E_{i}$ indicated in Fig.~4. 

We find a well-defined peak structure with a maximum of the intensity at about 1.4~eV kinetic energy and a weaker
shoulder at higher energies. According to the chosen photon energies for the pump and for the probe pulse, the
main peak is easily identified as the 2PPE signal from the first image state. The weak shoulder can be assigned
to the second and third image state. In this way, besides the fact that the experimental and theoretical time
scales are quite different, a direct comparison between our model calculations and corresponding experimental
data, as for example published by Shumay et al., \cite{SHR+98} is feasible.

\section{Summary}
We have presented a theoretical frame for the description and analysis of two-photon photoemission experiments which
is derived from the general theory of pump-probe photoemission by assuming that the pump pulse is weak in intensity.
The key quantity to describe time-revolved photoemission is the lesser Green function. We have shown that this quantity
is theoretically and numerically accessible for surfaces of realistic materials. For effectively independent electrons,
this is achieved by expressing the lesser Green function in terms of retarded and advanced Green functions. The latter
are obtained from a standard Dyson equation, where the perturbation is given by the pump pulse and treated perturbatively
in lowest order and where the unperturbed Green function is computed within multiple-scattering theory in a fully
relativistic way by using the Munich SPRKKR program package in its tight-binding version. As a first test case, the
theory has been applied to the (100) surface of Ag as a prototypical example of a simple metal. Our test calculations
based on a new numerical implementation clearly demonstrate the numerical feasibility of quantitative time-resolved
spectroscopic analysis for real materials.

\begin{acknowledgments}
Financial support by the Deutsche Forschungsgemeinschaft within research group FOR 1346, (projects P3 and P1),
within the collaborative research center SFB 925 (project B5), and projects Eb-154/23, Eb-154/26 as well as
by the BMBF project 05K13WMA is gratefully acknowledged. The work has also benefitted substantially from
discussions within the COST Action MP 1306 EUSpec.
\end{acknowledgments}

\bibliographystyle{aipnum}

\bibliography{akhelit}

\begin{thebibliography}{10}

\bibitem{LLM+05}
M.~Lisowski et~al.,
\newblock Phys. Rev. Lett. {\bf 95}, 137402 (2005).

\bibitem{CMU+07}
A.~L. Cavalieri et~al.,
\newblock Nature {\bf 449}, 1029 (2007).

\bibitem{PFB+08}
A.~Pietzsch et~al.,
\newblock New Journal of Physics {\bf 10}, 033004 (2008).

\bibitem{BPW10}
{\em Dynamics at Solid State Surfaces and Interfaces: Current Developments},
\newblock Wiley, New York, 2010.

\bibitem{RHW+11}
T.~Rohwer et~al.,
\newblock Nature {\bf 471}, 490 (2011).

\bibitem{GDP+11}
A.~Goris et~al.,
\newblock Phys. Rev. Lett. {\bf 107}, 026601 (2011).

\bibitem{GPM+11}
C.~M. G\"unther et~al.,
\newblock Nature Photonics {\bf 5}, 99 (2011).

\bibitem{VTT+12}
C.~La-O-Vorakiat et~al.,
\newblock Phys. Rev. X {\bf 2}, 011005 (2012).

\bibitem{CDF+12}
R.~Carley et~al.,
\newblock Phys. Rev. Lett. {\bf 109}, 057401 (2012).

\bibitem{SYA+12}
J.~A. Sobota et~al.,
\newblock Phys. Rev. Lett. {\bf 108}, 117403 (2012).

\bibitem{RVB+12}
D.~Rudolf et~al.,
\newblock Nature Communications {\bf 3}, 1037 (2012).

\bibitem{AGJH12}
N.~Armbrust, J.~G\"udde, P.~Jakob, and U.~H\"ofer,
\newblock Phys. Rev. Lett. {\bf 108}, 056801 (2012).

\bibitem{WHS+12}
Y.~H. Wang et~al.,
\newblock Phys. Rev. Lett. {\bf 109}, 127401 (2012).

\bibitem{DAB+13}
M.~Dell'Angela et~al.,
\newblock Science {\bf 339}, 1302 (2013).

\bibitem{MBV+13}
B.~Y. Mueller et~al.,
\newblock Phys. Rev. Lett. {\bf 111}, 167204 (2013).

\bibitem{AZ09}
M.~Aeschlimann and H.~Zacharias,
\newblock {\em Time-Resolved Two-Photon Photoemission on Surfaces and
  Nanoparticles}, volume~6 of {\em Nanotechnology}, page 273,
\newblock WILEY-VCH, Weinheim, 2009.

\bibitem{WSPD10}
M.~Weinelt, A.~B. Schmidt, M.~Pickel, and M.~Donath,
\newblock {\em Dynamics at Solid State Surfaces and Interfaces, Vol. 1: Current
  Developments}, chapter Spin-dependent relaxation of photoexcited electrons at
  surfaces of 3d ferromagnets,
\newblock Wiley, New York, 2010.

\bibitem{Fau12}
T.~Fauster,
\newblock {\em Surface and Interface Science}, chapter Two-photon photoelectron
  spectroscopy,
\newblock Wiley, Wiley-VCH (Weinheim), 2010.

\bibitem{RGK+14}
J.~Reimann, J.~G\"udde, K.~Kuroda, E.~V. Chulkov, and U.~H\"ofer,
\newblock Phys. Rev. B {\bf 90}, 081106 (2014).

\bibitem{BMA15}
M.~Bauer, A.~Marienfeld, and M.~Aeschlimann,
\newblock Progress in Surface Science {\bf 90}, 313 (2015).

\bibitem{KRGH16}
K.~Kuroda, J.~Reimann, J.~G\"udde, and U.~H\"ofer,
\newblock Phys. Rev. Lett. {\bf 116}, 076801 (2016).

\bibitem{CAH+06}
M.~Cinchetti et~al.,
\newblock Phys. Rev. Lett. {\bf 97}, 177201 (2006).

\bibitem{LLB+07}
P.~A. Loukakos et~al.,
\newblock Phys. Rev. Lett. {\bf 98}, 097401 (2007).

\bibitem{PSG+08}
M.~Pickel et~al.,
\newblock Phys. Rev. Lett. {\bf 101}, 066402 (2008).

\bibitem{SPD+10}
A.~B. Schmidt et~al.,
\newblock Phys. Rev. Lett. {\bf 105}, 197401 (2010).

\bibitem{SPA+08}
A.~B. Schmidt et~al.,
\newblock Journal of Physics D: Applied Physics {\bf 41}, 164003 (2008).

\bibitem{MPL+08}
A.~Melnikov et~al.,
\newblock Phys. Rev. Lett. {\bf 100}, 107202 (2008).

\bibitem{SGV+16}
J.~S\'anchez-Barriga et~al.,
\newblock Phys. Rev. B {\bf 93}, 155426 (2016).

\bibitem{WSMD07}
M.~Weinelt, A.~B. Schmidt, M.~Pickel, and M.~Donath,
\newblock Progress in Surface Science {\bf 82}, 388  (2007).

\bibitem{GRM+07}
J.~G\"udde, M.~Rohleder, T.~Meier, S.~W. Koch, and U.~H\"ofer,
\newblock Science {\bf 318}, 1287 (2007).

\bibitem{PSWD10}
M.~Pickel, A.~B. Schmidt, M.~Weinelt, and M.~Donath,
\newblock Phys. Rev. Lett. {\bf 104}, 237204 (2010).

\bibitem{Sch07}
A.~B. Schmidt,
\newblock {\em Spin-dependent electron dynamics infront of ferromagnetic
  surfaces},
\newblock PhD thesis, Freien Universit\"at Berlin, 2007.

\bibitem{MSS+11}
M.~Marks, C.~H. Schwalb, K.~Schubert, J.~G\"udde, and U.~H\"ofer,
\newblock Phys. Rev. B {\bf 84}, 245402 (2011).

\bibitem{FKP09}
J.~K. Freericks, H.~R. Krishnamurthy, and T.~Pruschke,
\newblock Phys. Rev. Lett. {\bf 102}, 136401 (2009).

\bibitem{SL13}
G.~Stefanucci and R.~van Leuween,
\newblock {\em Nonequilibrium Many-Body Theory of Quantum Systems},
\newblock Cambridge University Press, 2013.

\bibitem{SKM+13}
M.~Sentef et~al.,
\newblock Phys. Rev. X {\bf 3}, 041033 (2013).

\bibitem{USL14}
A.-M. Uimonen, G.~Stefanucci, and R.~van Leeuwen,
\newblock J. Phys. C: Solid State Phys. {\bf 140}, 18A526 (2014).

\bibitem{KSM+14}
A.~F. Kemper, M.~A. Sentef, B.~Moritz, J.~K. Freericks, and T.~P. Devereaux,
\newblock Phys. Rev. B {\bf 90}, 075126 (2014).

\bibitem{FNF14}
J.~K. Freericks, B.~N. Nikolic, and O.~Frieder,
\newblock International Journal of Modern Physics B {\bf 28}, 1430021 (2014).

\bibitem{SCK+15}
M.~A. Sentef et~al.,
\newblock Nature Communications {\bf 6}, 7047 (2015).

\bibitem{Bon16}
M.~Bonitz,
\newblock {\em Quantum Kinetic Theory},
\newblock Springer, Switzerland, 2016.

\bibitem{GKKR96}
A.~Georges, G.~Kotliar, W.~Krauth, and M.~J. Rozenberg,
\newblock Rev. Mod. Phys. {\bf 68}, 13 (1996).

\bibitem{KSH+06}
G.~Kotliar et~al.,
\newblock Rev. Mod. Phys. {\bf 78}, 865 (2006).

\bibitem{Hel07}
K.~Held,
\newblock Adv. Phys. {\bf 56}, 829 (2007).

\bibitem{KDE+15}
K.~Krieger, J.~K. Dewhurst, P.~Elliott, S.~Sharma, and E.~K.~U. Gross,
\newblock Journal of Chemical Theory and Computation {\bf 11}, 4870 (2015).

\bibitem{SHR+98}
V.~S. Stepanyuk et~al.,
\newblock Phys. Rev. B {\bf 57}, R14020 (1998).

\bibitem{NF14}
D.~Niesner and T.~Fauster,
\newblock Journal of Physics: Condensed Matter {\bf 26}, 093001 (2014).

\bibitem{NOH+14}
D.~Niesner et~al.,
\newblock Phys. Rev. B {\bf 89}, 081404 (2014).

\bibitem{GMTH14}
M.~C.~E. Galbraith, M.~Marks, R.~Tonner, , and U.~H\"ofer,
\newblock Journal of Physical Chemistry Letters {\bf 5}, 50 (2014).

\bibitem{BRP+15}
J.~Braun, R.~Rausch, M.~Potthoff, J.~Min\'ar, and H.~Ebert,
\newblock Phys. Rev. B {\bf 91}, 035119 (2015).

\bibitem{Kel65}
L.~V. Keldysh,
\newblock Sov. Phys. J.E.T.P. {\bf 20}, 1018 (1965).

\bibitem{MCP+05}
J.~Min\'ar et~al.,
\newblock Phys. Rev. B {\bf 72}, 045125 (2005).

\bibitem{BME+06}
J.~Braun, J.~Min\'ar, H.~Ebert, M.~I. Katsnelson, and A.~I. Lichtenstein,
\newblock Phys. Rev. Lett. {\bf 97}, 227601 (2006).

\bibitem{SFB+09}
J.~S\'anchez-Barriga et~al.,
\newblock Phys. Rev. Lett. {\bf 103}, 267203 (2009).

\bibitem{BMM+10}
J.~Braun, J.~Min\'ar, F.~Matthes, C.~M. Schneider, and H.~Ebert,
\newblock Phys. Rev. B {\bf 82}, 024411 (2010).

\bibitem{MBE13}
J.~Min\'ar, J.~Braun, and H.~Ebert,
\newblock J. Electron. Spectrosc. Relat. Phenom. {\bf 189}, 129 (2013).

\bibitem{Pen74}
J.~B. Pendry,
\newblock {\em Low energy electron diffraction},
\newblock Academic Press, London, 1974.

\bibitem{Pen76}
J.~B. Pendry,
\newblock Surf. Sci. {\bf 57}, 679 (1976).

\bibitem{HPT80}
J.~F.~L. Hopkinson, J.~B. Pendry, and D.~J. Titterington,
\newblock Comp. Phys. Commun. {\bf 19}, 69 (1980).

\bibitem{HK64}
P.~Hohenberg and W.~Kohn,
\newblock Phys. Rev. {\bf 136}, B 864 (1964).

\bibitem{GPU+11}
A.~X. Gray et~al.,
\newblock Nature Materials {\bf 10}, 759 (2011).

\bibitem{GMU+12}
A.~X. Gray et~al.,
\newblock Nature Materials {\bf 11}, 957 (2012).

\bibitem{MBE13a}
J.~Min\'ar, J.~Braun, and H.~Ebert,
\newblock J. Electron. Spectrosc. Relat. Phenom. {\bf 190}, 159 (2013).

\bibitem{BMK+14}
J.~Braun et~al.,
\newblock New Journal of Physics {\bf 16}, 015005 (2014).

\bibitem{BMM+13}
J.~Braun et~al.,
\newblock Phys. Rev. B {\bf 88}, 205409 (2013).

\bibitem{MDF10}
B.~Moritz, T.~P. Devereaux, and J.~K. Freericks,
\newblock Phys. Rev. B {\bf 81}, 165112 (2010).

\bibitem{EK08}
M.~Eckstein and M.~Kollar,
\newblock Phys. Rev. B {\bf 78}, 245113 (2008).

\bibitem{RFE16}
F.~{Randi}, D.~{Fausti}, and M.~{Eckstein},
\newblock ArXiv e-prints  (2016).

\bibitem{Ing11}
J.~E. Inglesfield,
\newblock J. Phys.: Cond. Mat. {\bf 23}, 305004 (2011).

\bibitem{UG07}
H.~Ueba and B.~Gumhalter,
\newblock Prog. Surr. Sci. {\bf 82}, 193 (2007).

\bibitem{Kor47}
J.~Korringa,
\newblock Physica {\bf 13}, 392 (1947).

\bibitem{EKM11}
H.~Ebert, D.~K\"odderitzsch, and J.~Min\'{a}r,
\newblock Rep. Prog. Phys. {\bf 74}, 096501 (2011).

\bibitem{Bra96}
J.~Braun,
\newblock Rep. Prog. Phys. {\bf 59}, 1267 (1996).

\bibitem{MCVP89}
J.~M. MacLaren, S.~Crampin, D.~D. Vvedensky, and J.~B. Pendry,
\newblock Phys. Rev. B {\bf 40}, 12164 (1989).

\bibitem{EBKM16}
H.~Ebert, J.~Braun, D.~K\"odderitzsch, and S.~Mankovsky,
\newblock Phys. Rev. B {\bf 93}, 075145 (2016).

\bibitem{GBB+93}
M.~Grass et~al.,
\newblock J. Phys. Condens. Matter {\bf 5}, 599 (1993).

\bibitem{MR80}
G.~Malmstr{\"o}m and J.~Rundgren,
\newblock Comp. Phys. Commun. {\bf 19}, 263  (1980).

\bibitem{SPR-KKR6.3}
{H.\ Ebert et al.},
\newblock {\em The Munich SPR-KKR package}, version 6.3, \newline
  \mbox{H.~Ebert~et~al.} \newline
  http://olymp.cup.uni-muenchen.de/ak/ebert/SPRKKR, 2012.

\bibitem{SFFS92}
S.~Schuppler, N.~Fischer, T.~Fauster, and W.~Steinmann,
\newblock Phys. Rev. B {\bf 46}, 13539 (1992).

\bibitem{Pic07}
M.~Pickel,
\newblock {\em Image-potential states as a sensor for magnetism},
\newblock PhD thesis, Freien Universit\"at Berlin, 2007.

\end{thebibliography}

\end{document}